\documentclass[a4paper,oldversion,letter]{aa}
\usepackage{graphics,txfonts}

\usepackage{natbib}
\bibpunct[, ]{(}{)}{;}{a}{}{,}

\newcommand{\teff}{$T_{\rm eff}$}
\newcommand{\lgg}{$\log g$}
\newcommand{\bz}{$\langle B_z\rangle$}
\newcommand{\vs}{$v_{\rm e}\sin i$}

\newcommand{\kms}{km\,s$^{-1}$}
\newcommand{\ms}{m\,s$^{-1}$}
\newcommand{\hd}{HD\,115226}
\newcommand{\pr}{\ion{Pr}{iii}}
\newcommand{\nd}{\ion{Nd}{iii}}
\newcommand{\dy}{\ion{Dy}{iii}}

\newcommand{\fifps}[2]{\centering\resizebox{#1}{!}{\includegraphics{#2}}}

\newcommand{\beq}{\begin{equation}}
\newcommand{\eeq}{\end{equation}}

\begin{document}

\title{The discovery of high-amplitude 10.9-minute oscillations \\ in the cool
magnetic Ap star HD\,115226%
\thanks{Based on observations collected at the European Southern Observatory,
La Silla, Chile (ESO program 079.D-0118)}}

\author{O. Kochukhov   \inst{1} \and
        T. Ryabchikova \inst{2,3} \and 
        S. Bagnulo     \inst{4} \and
        G. Lo Curto    \inst{5}}

\offprints{O. Kochukhov, \email{oleg@astro.uu.se}}

\institute{Department of Astronomy and Space Physics, Uppsala University, SE-751 20, Uppsala, Sweden
      \and Institute of Astronomy, Russian Academy of Sciences, Pyatnitskaya 48, 109017 Moscow, Russia
      \and Department of Astronomy, University of Vienna, T\"urkenschanzstra{\ss}e 17, 1180 Vienna, Austria
      \and Armagh Observatory, College Hill, Armagh BT61 9DG, Northern Ireland
      \and European Southern Observatory, Alonso de Cordova 3107, 
           Vitacura, Santiago, Casilla 19001 Santiago 19, Chile}

\date{Received / Accepted }

\abstract{
We present the discovery of pulsational variations in the cool magnetic Ap star
HD\,115226 -- the first high-amplitude rapidly oscillating Ap star discovered with
time-series spectroscopy. Using high-resolution spectra obtained with the HARPS
instrument at the ESO 3.6-m telescope, we detect radial velocity variations with a period
of 10.86~min in \ion{Pr}{iii}, \ion{Nd}{iii}, \ion{Dy}{iii} lines and in the narrow cores of hydrogen
lines. Pulsational amplitudes exceed 1~\kms\ in individual lines of \ion{Nd}{iii}. The
presence of running waves in the stellar atmosphere is inferred from a phase shift
between the radial velocity maxima of rare-earth and hydrogen lines. Our abundance
analysis demonstrates that HD\,115226 exhibits typical roAp spectroscopic signature,
notably ionization anomaly of Pr, Nd and Dy. We discuss the discovery of
pulsations in HD\,115226 in the context of recent spectroscopic studies of roAp stars and
point to the existence of correlation between spectroscopic pulsational amplitude and 
the stellar rotation rate.
}


\maketitle

\section{Introduction}
\label{intro}

Rapidly oscillating Ap (roAp) stars are cool magnetic chemically peculiar stars, pulsating in high-overtone
non-radial {\it p-}modes with periods around 10~min. Excitation of oscillations in these stars is closely
related to the presence of strong global magnetic field, as the pulsational axis is seen to be aligned with the axis of
oblique magnetic field and pulsations are enhanced at the magnetic poles \citep{kochukhov:2004vx}. In addition
to remarkable pulsation behaviour, which opens interesting prospects for asteroseismology 
\citep{vauclair:2004lq}, roAp stars are characterized by extremely anomalous surface chemical composition 
\citep{ryabchikova:2004uc}, as well as strongly inhomogeneous horizontal and vertical distribution of chemical
elements \citep{kochukhov:2004vp,ryabchikova:2002rk}.

Most of the previous work on identification and frequency analysis of the three dozen currently known roAp stars
was performed using time-series photometry at small telescopes \citep{kurtz:2000oj}. However, inspired by the
discovery of outstanding pulsational variations in the lines of rare-earth elements (REEs) 
\citep{savanov:1999fj,kochukhov:2001tm}, the focus of observational studies of roAp stars is shifting towards
time-resolved spectroscopy. A number of recent studies \citep[e.g.,][]{mkrtichian:2003gk,ryabchikova:2007lq}
have demonstrated unique diagnostic potential of the precise radial velocity and line profile measurements in
individual lines. This type of observational material allows one to trace propagation of pulsation waves through
the chemically stratified stellar atmosphere, eventually obtaining a three-dimensional picture of 
pulsations and chemical inhomogeneities \citep{kochukhov:2007fp}.

Despite these noteworthy achievements of time-resolved spectroscopic studies, relatively little is done in the
direction of expanding the list of known roAp stars and understanding the coexistence of pulsating and apparently
constant Ap stars in the same region of the H-R diagramm. With the aim to obtain an improved picture of
the incidence of pulsations in Ap stars, we have initiated a survey of cool peculiar magnetic stars with the
ultra-stable spectrograph HARPS at ESO. Here we report the first result of our observations -- the discovery of
high-amplitude oscillations in the magnetic Ap star \hd.

\section{Observations and data reduction}
\label{obs}

We have observed \hd\ on the night of April 15, 2007, with the HARPS spectrograph
\citep{mayor:2003qy} at the ESO 3.6-m telescope at La Silla. The observations have
started at the barycentric JD 2454205.63472 and continued for 4.3 hours. In total, 102
consecutive 120~s exposures were collected. The dead time between stellar observations
was 31~s.

The extraction of one-dimensional spectra and barycentric velocity correction of the
wavelength scale was carried out with the HARPS pipeline available at the telescope. 
The peak signal-to-noise ratio of 40--45 was achieved in the individual exposures of \hd.
The spectra have resolving power $\lambda/\Delta\lambda=115\,000$ and cover a 
wavelength range from 3780 to 6910~\AA, with a 30~\AA\ gap near 5320~\AA. 

The dominant source of radial velocity noise in our data will be due to the
photon noise rather than to the instrumental precision. Therefore, we
did use the simultaneous ThAr method, avoiding in this way
any contamination of the stellar signal. At the beginning and at the end
of \hd\ observations we took a ThAr reference spectrum, and can estimate that the
instrumental drift within the time series was at most 0.36~\ms.

Following the procedure outlined in \citet{kochukhov:2007db}, we post-processed all
extracted spectra of \hd\ with the aim to achieve an accurate and consistent continuum
normalization.

\section{Basic properties of HD\,115226}
\label{atmos}

Little is known about \hd\ (HIP~64883, CP$-$72~1373). This southern $V=8.51$ chemically peculiar star is
classified as A3p~Sr in the General Catalogue of Ap and Am Stars \citep{renson:1991tl}.
\citet{gomez:1998yl} and \citet{kochukhov:2006dj} included this object in their studies of 
the evolutionary state of
magnetic Ap stars. Using Hipparcos parallax and \teff\,=\,7640~K estimated from the
Geneva photometric colors, the authors of the latter paper determined $L$\,=\,$7.2\pm1.9$~$L_\odot$. From the
comparison with standard stellar evolutionary models \citep{schaller:1992by} one can 
infer that these parameters correspond to a $M$\,=\,$1.60\pm0.05$~$M_\odot$ star with a
fractional age between 0.0 and 0.4 of the main sequence lifetime.

The Str\"omgren photometry of \hd\ \citep{martinez:1993qy} indicates \teff\,=\,8000--8200~K if
normal-star calibrations of \citet{moon:1985xz} and \citet{napiwotzki:1993uj} are used. However,
comparison of the theoretical calculations with the observed profiles of H$\alpha$ and H$\beta$ 
favours lower \teff\ suggested by the Geneva photometric parameters.

\begin{figure}[!th]
\fifps{8cm}{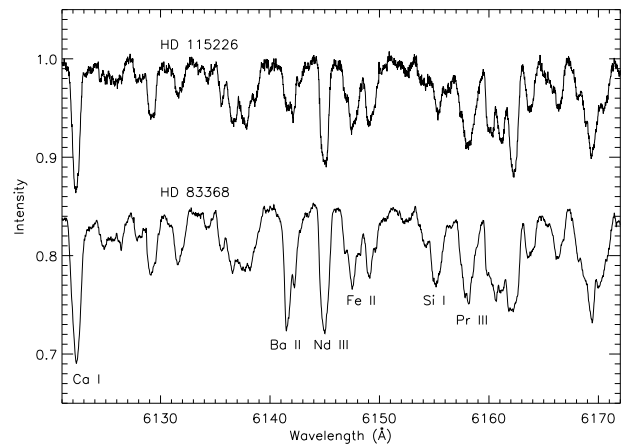}
\caption{Comparison of the 6120--6170~\AA\ region in the average spectrum of \hd\ with the 
known rapidly rotating roAp star HD\,83368 \citep{kochukhov:2004vp}. Prominent lines of \pr\
and \nd\ appear in both stars.
The spectrum of HD\,83368 is shifted downwards for display purpose.}
\label{fig0}
\end{figure}

\citet{kochukhov:2006dj} determined longitudinal field \bz\,=\,$752\pm48$~G from a single circular
polarization measurement of \hd\ with the FORS1 instrument at VLT. This robust magnetic field
detection, along with the typical peculiar-star appearance of the spectrum of \hd\
(Fig.~\ref{fig0}), leaves no doubt that this star belongs to the group of cool magnetic Ap stars.
The value of \teff\ inferred for \hd\ is also typical of the known roAp stars.

To perform a preliminary analysis of the atmospheric chemistry of \hd\ and to compare it with the
chemical abundances of well-studied roAp stars, we adopted a model atmosphere of the roAp star
10~Aql  (HD~176232) with the parameters \teff\,=\,7650 K, \lgg\,=4.0, determined in our study of Ca
stratification in 10~Aql \citep{ryabchikova:2007fj}. A rather rapid rotation of \hd\ does  not
allow us to study individual lines for most elements, therefore we performed magnetic spectral 
synthesis with the help of {\sc synthmag} code \citep{kochukhov:2007vn}. First of all, we tried to
estimate \vs\ and the strength of surface magnetic field using magnetic intensification. Line
profiles of almost all elements in the spectrum of \hd\ have very complex structure, indicating a
high degree of the horizontal abundance inhomogeneity. Best-fit projected rotational velocities
vary from 18 \kms\ for \ion{Co}{i} and \ion{Y}{ii} to 23--25 \kms\ for REEs and \ion{Si}{ii} lines
and up to 28--30 \kms\ for Mg, Ca, Cr and Fe lines. This dispersion in broadening is accompanied by
different shifts of the line centroids: $-10$~\kms\ for \ion{Co}{i}, from $-5$ to $-8$ \kms\ in
REEs, from $-2$  to $-3$~\kms\ in Mg, Ca, Cr, Fe, and $+7$ \kms\ for Si. These discrepancies
point to a spotted distribution of chemical elements at the stellar surface.

Our spectrum synthesis was performed for the magnetic field strength in the range
from 1 to 3~kG. The lowest dispersion in the abundances inferred from spectral lines with different
Land\'e factors (0--2.5 for Fe lines and 0.6--1.6 for \ion{Nd}{iii} lines) and different 
Zeeman structure was obtained for the magnetic field strength of 1~kG.

The result of abundance analysis of \hd\ is summarized in Table\,\ref{tbl0}.
The inferred abundances are typical of roAp stars in the 7500--8000~K effective
temperature range. The star has nearly solar concentration of Mg, Si, Ti, Fe, overabundance of Cr
and Co, underabundance of Ba. The well-known REE anomaly, where abundance derived from the lines
of second ions (\ion{Pr}{iii}, \ion{Nd}{iii}, \ion{Dy}{iii}) exceeds by 1.5--2~dex those derived
from the lines of singly ionized species \citep{ryabchikova:2001ub,ryabchikova:2004uc}, is also present.
Interestingly, the vertical Ca abundance distribution determined for the atmosphere of 10~Aql fits
well the Ca line  profiles in the spectrum of \hd. This suggests close similarity of the Ca
stratifications in the atmospheres of these two stars.

\begin{table}
\centering
\caption{Chemical abundances in the atmosphere of \hd. 
Error estimates are given for species with
sufficient number of measured lines. 
In other cases we provide a range of abundance that fits
individual spectral features. \label{tbl0}}
\begin{tabular}{ll|ll}
\hline
\hline
Ion & $\log(N_{\rm el}/N_{\rm tot})$ & Ion & $\log(N_{\rm el}/N_{\rm tot})$ \\
\hline
\ion{Mg}{i}/{\sc ii}    & ~$-$4.5           & \ion{Ce}{ii}  & ~$-$9.5	\\                   
\ion{Si}{i}/{\sc ii}    & ~$-$4.6           & \ion{Pr}{ii}  & ~$-$9.6	\\           
\ion{Ca}{i}/{\sc ii}    & ~$-5.4\pm0.2$     & \ion{Pr}{iii} & ~$-8.1\pm0.2$	\\
\ion{Ti}{ii}            & ~$-$7.0           & \ion{Nd}{ii}  & ~$-$8.8 {\rm to} $-$9.0		\\           
\ion{Cr}{i}/{\sc ii}    & ~$-$5.0           & \ion{Nd}{iii} & ~$-7.2\pm0.3$	\\        
\ion{Fe}{i}/{\sc ii}    & ~$-$4.5 {\rm to} $-$4.6 & \ion{Sm}{ii}  & ~$-$9.2 {\rm to} $-$9.5	\\ 
\ion{Ni}{i}             & ~$-$6.3           & \ion{Eu}{ii}  & $-$10.2     \\          
\ion{Co}{i}             & ~$-$6.0           & \ion{Gd}{ii}  & ~$-$9.3     \\
\ion{Y}{ii}             & ~$-$8.5           & \ion{Tb}{iii} & ~$-$8.2     \\           
\ion{Ba}{ii}            & $-$10.5           & \ion{Dy}{ii}  & ~$-$9.0     \\           
\ion{La}{ii}            & ~$-$9.50          & \ion{Dy}{iii} & ~$-7.3\pm0.2$       \\        
\hline                                                                   
\end{tabular}
\end{table}

Significant Doppler broadening of the \hd\ spectrum indicates that the stellar
rotational period is short. Combining \vs\,=\,25--30~\kms\ determined here with the
information on stellar luminosity and temperature, we find $P_{\rm rot}$\,$\le$\,3.0--3.5~d. 
However, despite the presence of surface abundance
inhomogeneities, \hd\ does not exhibit a prominent variability in the 
ASAS \citep{pojmanski:2002lq} photometric time-series. At the same time, Hipparcos
Epoch Photometry of this star \citep{esa:1997qr} shows
a marginal variation with $P=3.61$~d and an amplitude below 0.01~mag.

\section{Analysis of pulsational variability}
\label{puls}

Large rotational broadening in the spectrum of \hd\ significantly reduces the number of
lines suitable for precise radial velocity measurements. Fortunately, the star shows
strong lines of \pr\ and \nd, which exhibit pulsational variability in almost all known
roAp stars \citep{kochukhov:2001tm,ryabchikova:2007lq}. Using the VALD database 
\citep{kupka:1999hp} and an improved \nd\ line list published by
\citet{ryabchikova:2006yu}, we measured radial velocities of the absorption features of
doubly ionized REEs and the hydrogen line cores. Velocities were
determined with the centre-of-gravity method. Based on these measurements, the amplitudes
and phases of variable lines were found by least-squares fitting of the cosine function
$A\cos{\left[2\pi(t/P+\varphi)\right]}$.

The radial velocity analysis of strongest rare-earth lines readily shows pulsational
signal with the frequency $\approx$\,1.53~mHz. The cores of H$\alpha$ and H$\beta$, as
well as weaker lines of doubly ionized Pr, Nd and Dy, also
show variation with this frequency. Thus, our observations demonstrate that \hd\ is a new roAp
star, pulsating with a period typical of this class of oscillating stars. However, the
radial velocity amplitude in \hd\ significantly exceeds the average level of few hundred
\ms\ seen in other roAp stars.

\begin{figure}[!th]
\fifps{8cm}{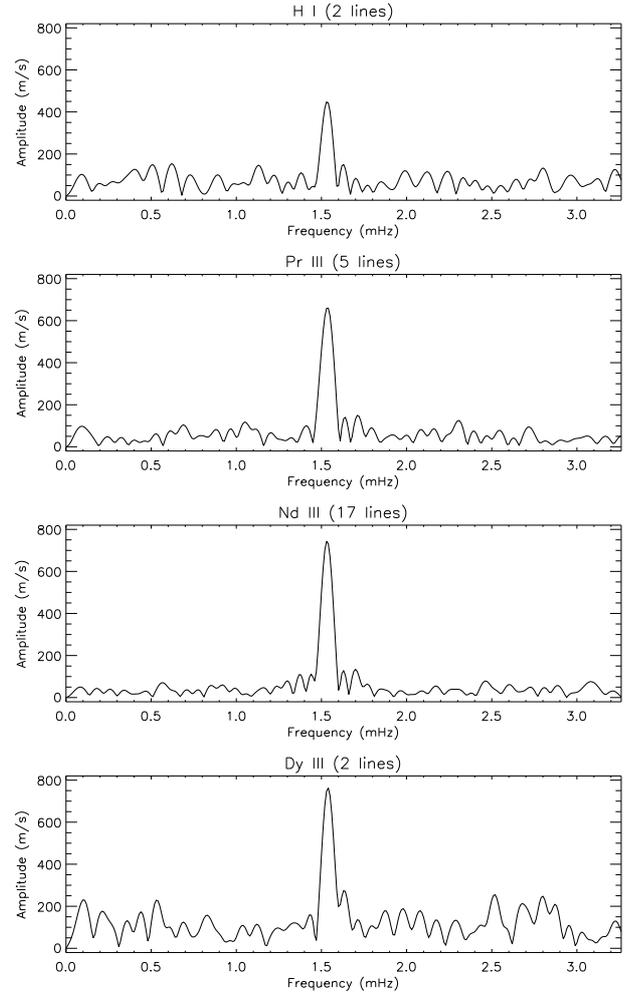}
\caption{Amplitude spectra for the combined radial velocity of two \ion{H}{i}  
5 \pr, 17 \nd, and two \dy\ lines.
The presence of rapid oscillation with the frequency 1.53~mHz is
evident.}
\label{fig1}
\end{figure}

\begin{figure}[!th]
\fifps{8cm}{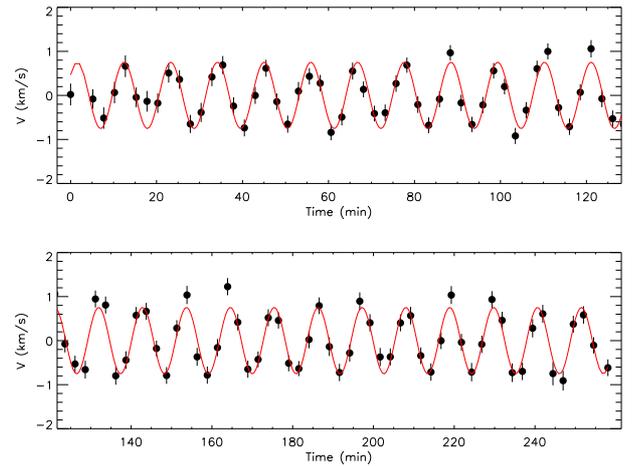}
\caption{Average radial velocity of 17 \nd\ lines as a function of time. Symbols show
observations, the solid curve illustrates a least-squares cosine fit with $P=10.864$~min.}
\label{fig2}
\end{figure}

In order to determine pulsational characteristics of \hd\ with a better precision, we
analysed combined radial velocity measurements of \ion{H}{i} (2 lines), \pr\ (5 lines) and
\nd\ (17 lines). Corresponding amplitude spectra are presented in Fig.~\ref{fig1}. 
Combined velocity of the two weak \ion{Dy}{iii} lines shows higher noise level 
(lower panel in Fig.~\ref{fig1}), therefore we disregard this REE ion in the period determination. 
The least-squares fit yields $P=10.880\pm0.031$, $10.852\pm0.016$ and $10.867\pm0.010$~min for
\ion{H}{i}, \pr\ and \nd, respectively. The weighted mean value of the pulsation period is
thus $10.864\pm0.008$~min (frequency 1.534~mHz). Fig.~\ref{fig2} shows the average radial
velocity curve of \nd\ together with a sinusoid fit. Analysis of the residuals shows no
evidence of additional frequencies with amplitudes above 110~\ms. For \ion{H}{i} and \pr\
the corresponding noise level (amplitude of the highest noise peak in the residuals)
is 160~\ms. Further spectroscopic observations at a larger telescope are required to
confirm monoperiodic character of oscillations in \hd.

\begin{table}[!th]
\centering
\caption{Amplitudes and phases of individual lines of \ion{H}{i}, \pr, \nd\ and \dy\
determined by the least-squares analysis of radial velocity measurements. 
The phases, determined with respect to the BJD of the first observation (see Sect.~\ref{obs}), 
are given in units of pulsation period.
\label{tbl1}}
\begin{tabular}{lcrr}
\hline
\hline
Ion        & $\lambda$ (\AA) & $A$ (\ms) & $\varphi$~~~~~~~~~ \\
\hline
\ion{H}{i} & 4861.32 & $ 423\pm\phantom{1}81$ & $0.043\pm0.033$ \\
\ion{H}{i} & 6562.78 & $ 526\pm\phantom{1}58$ & $0.021\pm0.019$ \\
\hline                                               
\pr\       & 5956.04 & $ 777\pm\phantom{1}94$ & $0.820\pm0.021$ \\
\pr\       & 6090.01 & $ 644\pm\phantom{1}86$ & $0.837\pm0.023$ \\
\pr\       & 6160.23 & $ 841\pm108$ & $0.795\pm0.022$ \\
\pr\       & 6195.62 & $ 622\pm\phantom{1}72$ & $0.764\pm0.020$ \\
\pr\       & 6866.79 & $ 886\pm\phantom{1}90$ & $0.871\pm0.017$ \\
\hline                                               
\nd\       & 4759.53 & $ 616\pm117$ & $0.925\pm0.032$ \\
\nd\       & 4769.62 & $ 714\pm101$ & $0.906\pm0.024$ \\
\nd\       & 4911.65 & $ 620\pm\phantom{1}63$ & $0.885\pm0.017$ \\
\nd\       & 4914.09 & $1242\pm\phantom{1}97$ & $0.870\pm0.013$ \\
\nd\       & 4927.57 & $ 504\pm\phantom{1}53$ & $0.833\pm0.018$ \\
\nd\       & 4942.64 & $ 430\pm\phantom{1}66$ & $0.831\pm0.026$ \\
\nd\       & 5050.70 & $ 949\pm\phantom{1}50$ & $0.853\pm0.009$ \\
\nd\       & 5102.42 & $ 763\pm\phantom{1}62$ & $0.791\pm0.014$ \\
\nd\       & 5126.99 & $ 546\pm\phantom{1}55$ & $0.809\pm0.017$ \\
\nd\       & 5294.11 & $ 908\pm\phantom{1}96$ & $0.778\pm0.018$ \\
\nd\       & 5677.15 & $ 744\pm108$ & $0.863\pm0.025$ \\
\nd\       & 5802.53 & $1183\pm173$ & $0.899\pm0.025$ \\
\nd\       & 5845.07 & $ 554\pm\phantom{1}71$ & $0.824\pm0.021$ \\
\nd\       & 5851.54 & $1212\pm122$ & $0.872\pm0.017$ \\
\nd\       & 6145.07 & $ 766\pm122$ & $0.855\pm0.027$ \\
\nd\       & 6327.26 & $1011\pm108$ & $0.874\pm0.018$ \\
\nd\       & 6550.24 & $1005\pm198$ & $0.786\pm0.033$ \\
\hline
\dy\       & 4510.03 & $789\pm\phantom{1}82$ & $0.928\pm0.017$ \\
\dy\       & 5730.28 & $823\pm130$  & $0.956\pm0.027$ \\
\hline                                               
\end{tabular}
\end{table}

Finally, we have derived least-squares estimate of the amplitudes and phases of 26
variable lines for the fixed 10.864~min period (Table~\ref{tbl1}). Integration time of
120~s adopted in our observations of \hd\ is non-negligible compared to the inferred
pulsational period. We found that for the time sampling of our data the phase smearing
reduces amplitudes by 6.2~\%. Consequently, all radial velocity amplitudes reported
here are multiplied by a factor 1.066 to correct for this effect.

From Table~\ref{tbl1} it is evident that oscillation amplitudes show considerable scatter
even for the lines of the same ion. This can be attributed to the dilution of 
pulsation signal in the lines affected by unrecognized blends. For this reason
our amplitude determination possibly gives only a lower limit of the real
signal for many pulsating lines. Thus, we believe that the radial velocity amplitudes exceeding 1~\kms, as seen
for \nd\ 4914, 5802, 5851, 6327 and 6550~\AA, are more representative of the pulsations in 
\hd\ than variability in the lines with amplitudes around 500~\ms.

Time-series analysis reveals that the lines of \pr\ and \nd\ pulsate with approximately the
same phase, which differs from that of \dy\ and \ion{H}{i}. In our convention a larger
pulsation phase corresponds to an earlier radial velocity maximum. Therefore, we find
that the \dy\ velocity lags by 0.1 period behind \ion{H}{i}. The phase lag reaches 0.2 
for \pr\ and \nd.

A number of strong lines belonging to the iron-peak elements and Ca were also searched
for pulsation signatures. No variability exceeding 100~\ms\ was detected in these lines.
           
\section{Discussion}
\label{discus}

We show that \hd\ is a new roAp star with broad lines and prominent pulsational variability in the cores
of hydrogen lines and in doubly ionized Pr, Nd and Dy. The radial velocity amplitudes, exceeding 1~\kms\ in
some \nd\ lines, are rather high for a roAp star. On the other hand, atmospheric parameters and chemical
abundance pattern of \hd\ are very close to those of known roAp stars. The phase shifts between radial velocity
variation of different spectral lines suggest the presence of running pulsation wave, first propagating in the
atmospheric layers where hydrogen lines form and then reaching the heights where REEs are enhanced. This
behaviour is observed in many other roAp stars \citep{ryabchikova:2007lq}.

Recently pulsational radial velocity variations at the level below 100~\ms\ were reported for HD\,137909
\citep{hatzes:2004lq}, HD\,116114 \citep{elkin:2005fs}, HD\,154708 \citep{kurtz:2006mi} and HD\,75445
\citep{kochukhov:2007qy}. In contrast, our discovery of oscillations in \hd\ adds to the list of roAp stars with
much higher amplitudes.
In fact, given the low
probability of obtaining observations at the rotation phase of maximum pulsational variability, it is quite
possible that \hd\ has one of the highest radial velocity amplitudes among all roAp stars.

Rapid rotation and high pulsational amplitude makes \hd\ similar to HD\,83368 \citep{kochukhov:2006ts} and
HD\,99563 \citep{elkin:2005rp}. Like these two well-studied roAp stars, \hd\ is an obvious candidate for
monitoring pulsations over complete rotation cycle. We propose to obtain such observations both in photometry 
(to determine rotation period and to study the pulsation frequency spectrum) and in spectroscopy (to derive
distribution of chemical spots and to study the topology of pulsation velocity field).

The discovery of strong variability in \hd\ hints at the existence of a relationship between the stellar
rotation and excitation of pulsations. The group of roAp stars which show radial velocity amplitudes in \nd\ at
the level of 1~\kms\ and above now includes HD\,60435, HD\,83368, HD\,99563, HD\,115226, HD\,12932 and HD\,19918.
The first four stars all have \vs\,$>$\,10~\kms\ and $P_{\rm rot}$\,$<$\,8~d. Rotation periods are unknown for
the two latter stars, but cannot be long given \vs\,$=$\,3.0--3.5~\kms\ measured for these objects
\citep{ryabchikova:2007lq}. Thus, we come to the conclusion that \textit{the high-amplitude roAp stars are
markedly different in their rotation properties from roAp stars with lower pulsational amplitudes}, many
of which have rotation periods of the order of years and even decades. In the light of this finding, the problem
of the influence of rotation on excitation of {\it p-}modes in roAp stars deserves careful theoretical
consideration.

\begin{acknowledgements}
This work was supported by the RFBI grant 06-02-16110a, RAS Presidium Program ``Origin and Evolution 
of Stars and Galaxies'', and the Austrian Science Fund (FWF projects P17580, P17890).
\end{acknowledgements}


\begin{thebibliography}{33}
\expandafter\ifx\csname natexlab\endcsname\relax\def\natexlab#1{#1}\fi

\bibitem[{{Elkin} {et~al.}(2005{\natexlab{a}}){Elkin}, {Kurtz}, \&
  {Mathys}}]{elkin:2005rp}
{Elkin}, V.~G., {Kurtz}, D.~W., \& {Mathys}, G. 2005{\natexlab{a}}, \mnras,
  364, 864

\bibitem[{{Elkin} {et~al.}(2005{\natexlab{b}}){Elkin}, {Riley}, {Cunha},
  {Kurtz}, \& {Mathys}}]{elkin:2005fs}
{Elkin}, V.~G., {Riley}, J.~D., {Cunha}, M.~S., {Kurtz}, D.~W., \& {Mathys}, G.
  2005{\natexlab{b}}, \mnras, 358, 665

\bibitem[{{ESA}(1997)}]{esa:1997qr}
{ESA} 1997, ESA Special Publication, Vol. 1200, {The HIPPARCOS and TYCHO
  catalogues}

\bibitem[{{Gomez} {et~al.}(1998){Gomez}, {Luri}, {Grenier}, {Figueras},
  {North}, {Royer}, {Torra}, \& {Mennessier}}]{gomez:1998yl}
{Gomez}, A.~E., {Luri}, X., {Grenier}, S., {et~al.} 1998, \aap, 336, 953

\bibitem[{{Hatzes} \& {Mkrtichian}(2004)}]{hatzes:2004lq}
{Hatzes}, A.~P. \& {Mkrtichian}, D.~E. 2004, \mnras, 351, 663

\bibitem[{{Kochukhov}(2004)}]{kochukhov:2004vx}
{Kochukhov}, O. 2004, \apjl, 615, L149

\bibitem[{{Kochukhov}(2006)}]{kochukhov:2006ts}
{Kochukhov}, O. 2006, \aap, 446, 1051

\bibitem[{{Kochukhov}(2007{\natexlab{a}})}]{kochukhov:2007fp}
{Kochukhov}, O. 2007{\natexlab{a}}, Communications in Asteroseismology, 150, 39

\bibitem[{{Kochukhov}(2007{\natexlab{b}})}]{kochukhov:2007vn}
{Kochukhov}, O. 2007{\natexlab{b}}, in Physics of Magnetic Stars, eds. I.~I.
  {Romanyuk}, D.~O. {Kudryavtsev}, 109--118

\bibitem[{{Kochukhov} \& {Bagnulo}(2006)}]{kochukhov:2006dj}
{Kochukhov}, O. \& {Bagnulo}, S. 2006, \aap, 450, 763

\bibitem[{{Kochukhov} {et~al.}(2004){Kochukhov}, {Drake}, {Piskunov}, \& {de la
  Reza}}]{kochukhov:2004vp}
{Kochukhov}, O., {Drake}, N.~A., {Piskunov}, N., \& {de la Reza}, R. 2004,
  \aap, 424, 935

\bibitem[{{Kochukhov} \& {Ryabchikova}(2001)}]{kochukhov:2001tm}
{Kochukhov}, O. \& {Ryabchikova}, T. 2001, \aap, 374, 615

\bibitem[{{Kochukhov} {et~al.}(2007{\natexlab{a}}){Kochukhov}, {Ryabchikova},
  {Bagnulo}, \& {Lo Curto}}]{kochukhov:2007qy}
{Kochukhov}, O., {Ryabchikova}, T., {Bagnulo}, S., \& {Lo Curto}, G.
  2007{\natexlab{a}}, in CP\#Ap Workshop, 
eds. J.\,\v{Z}i\v{z}\v{n}ovsk\'{y}, J. Zverko, E. Paunzen, M. Netopil,  
Contribut. Astr. Obs. Skalnat\'e Pleso, in press (astro-ph/0711.4923)

\bibitem[{{Kochukhov} {et~al.}(2007{\natexlab{b}}){Kochukhov}, {Ryabchikova},
  {Weiss}, {Landstreet}, \& {Lyashko}}]{kochukhov:2007db}
{Kochukhov}, O., {Ryabchikova}, T., {Weiss}, W.~W., {Landstreet}, J.~D., \&
  {Lyashko}, D. 2007{\natexlab{b}}, \mnras, 376, 651

\bibitem[{{Kupka} {et~al.}(1999){Kupka}, {Piskunov}, {Ryabchikova}, {Stempels},
  \& {Weiss}}]{kupka:1999hp}
{Kupka}, F., {Piskunov}, N., {Ryabchikova}, T.~A., {Stempels}, H.~C., \&
  {Weiss}, W.~W. 1999, \aaps, 138, 119

\bibitem[{{Kurtz} {et~al.}(2006){Kurtz}, {Elkin}, {Cunha}, {Mathys}, {Hubrig},
  {Wolff}, \& {Savanov}}]{kurtz:2006mi}
{Kurtz}, D.~W., {Elkin}, V.~G., {Cunha}, M.~S., {et~al.} 2006, \mnras, 372, 286

\bibitem[{{Kurtz} \& {Martinez}(2000)}]{kurtz:2000oj}
{Kurtz}, D.~W. \& {Martinez}, P. 2000, Baltic Astronomy, 9, 253

\bibitem[{{Martinez}(1993)}]{martinez:1993qy}
{Martinez}, P. 1993, PhD thesis, , University of Cape Town, SA, (1993)

\bibitem[{{Mayor} {et~al.}(2003){Mayor}, {Pepe}, {Queloz}, {Bouchy},
  {Rupprecht}, {Lo Curto}, {Avila}, {Benz}, {Bertaux}, {Bonfils}, {dall},
  {Dekker}, {Delabre}, {Eckert}, {Fleury}, {Gilliotte}, {Gojak}, {Guzman},
  {Kohler}, {Lizon}, {Longinotti}, {Lovis}, {Megevand}, {Pasquini}, {Reyes},
  {Sivan}, {Sosnowska}, {Soto}, {Udry}, {van Kesteren}, {Weber}, \&
  {Weilenmann}}]{mayor:2003qy}
{Mayor}, M., {Pepe}, F., {Queloz}, D., {et~al.} 2003, The Messenger, 114, 20

\bibitem[{{Mkrtichian} {et~al.}(2003){Mkrtichian}, {Hatzes}, \&
  {Kanaan}}]{mkrtichian:2003gk}
{Mkrtichian}, D.~E., {Hatzes}, A.~P., \& {Kanaan}, A. 2003, \mnras, 345, 781

\bibitem[{{Moon} \& {Dworetsky}(1985)}]{moon:1985xz}
{Moon}, T.~T. \& {Dworetsky}, M.~M. 1985, \mnras, 217, 305

\bibitem[{{Napiwotzki} {et~al.}(1993){Napiwotzki}, {Schoenberner}, \&
  {Wenske}}]{napiwotzki:1993uj}
{Napiwotzki}, R., {Schoenberner}, D., \& {Wenske}, V. 1993, \aap, 268, 653

\bibitem[{{Pojmanski}(2002)}]{pojmanski:2002lq}
{Pojmanski}, G. 2002, Acta Astronomica, 52, 397

\bibitem[{{Renson} {et~al.}(1991){Renson}, {Gerbaldi}, \&
  {Catalano}}]{renson:1991tl}
{Renson}, P., {Gerbaldi}, M., \& {Catalano}, F.~A. 1991, \aaps, 89, 429

\bibitem[{{Ryabchikova} {et~al.}(2007{\natexlab{a}}){Ryabchikova}, {Kochukhov},
  \& {Bagnulo}}]{ryabchikova:2007fj}
{Ryabchikova}, T., {Kochukhov}, O., \& {Bagnulo}, S. 2007{\natexlab{a}}, in
  Physics of Magnetic Stars, eds. I.~I. {Romanyuk} \& D.~O. {Kudryavtsev},
  325--334

\bibitem[{{Ryabchikova} {et~al.}(2004){Ryabchikova}, {Nesvacil}, {Weiss},
  {Kochukhov}, \& {St{\"u}tz}}]{ryabchikova:2004uc}
{Ryabchikova}, T., {Nesvacil}, N., {Weiss}, W.~W., {Kochukhov}, O., \&
  {St{\"u}tz}, C. 2004, \aap, 423, 705

\bibitem[{{Ryabchikova} {et~al.}(2002){Ryabchikova}, {Piskunov}, {Kochukhov},
  {Tsymbal}, {Mittermayer}, \& {Weiss}}]{ryabchikova:2002rk}
{Ryabchikova}, T., {Piskunov}, N., {Kochukhov}, O., {et~al.} 2002, \aap, 384,
  545

\bibitem[{{Ryabchikova} {et~al.}(2006){Ryabchikova}, {Ryabtsev}, {Kochukhov},
  \& {Bagnulo}}]{ryabchikova:2006yu}
{Ryabchikova}, T., {Ryabtsev}, A., {Kochukhov}, O., \& {Bagnulo}, S. 2006,
  \aap, 456, 329

\bibitem[{{Ryabchikova} {et~al.}(2007{\natexlab{b}}){Ryabchikova}, {Sachkov},
  {Kochukhov}, \& {Lyashko}}]{ryabchikova:2007lq}
{Ryabchikova}, T., {Sachkov}, M., {Kochukhov}, O., \& {Lyashko}, D.
  2007{\natexlab{b}}, \aap, 473, 907

\bibitem[{{Ryabchikova} {et~al.}(2001){Ryabchikova}, {Savanov}, {Malanushenko},
  \& {Kudryavtsev}}]{ryabchikova:2001ub}
{Ryabchikova}, T.~A., {Savanov}, I.~S., {Malanushenko}, V.~P., \&
  {Kudryavtsev}, D.~O. 2001, Astronomy Reports, 45, 382

\bibitem[{{Savanov} {et~al.}(1999){Savanov}, {Malanushenko}, \&
  {Ryabchikova}}]{savanov:1999fj}
{Savanov}, I.~S., {Malanushenko}, V.~P., \& {Ryabchikova}, T.~A. 1999,
  Astronomy Letters, 25, 802

\bibitem[{{Schaller} {et~al.}(1992){Schaller}, {Schaerer}, {Meynet}, \&
  {Maeder}}]{schaller:1992by}
{Schaller}, G., {Schaerer}, D., {Meynet}, G., \& {Maeder}, A. 1992, \aaps, 96,
  269

\bibitem[{{Vauclair} \& {Th{\'e}ado}(2004)}]{vauclair:2004lq}
{Vauclair}, S. \& {Th{\'e}ado}, S. 2004, \aap, 425, 179

\end{thebibliography}

\end{document}